\documentclass{svjour3}                     
\smartqed  
\usepackage{graphicx}
\usepackage{graphicx}
\DeclareGraphicsExtensions{.eps,.png}
\usepackage{amsmath}
\usepackage{epsfig}
\usepackage{version}
 \usepackage{mathptmx}      
\begin{document}

\title{Generalized parton distributions~of $^3$He
and the neutron orbital structure
}


\author{M. Rinaldi         \and
        S. Scopetta 
}


\institute{M. Rinaldi \at
	Dipartimento di Fisica, Universit\`{a} degli studi di Perugia 
and INFN sezione di Perugia, Via A. Pascoli 06100 Perugia, Italy
 \\
\email{matteo.rinaldi@pg.infn.it}           
\and
S. Scopetta \at
Dipartimento di Fisica, Universit\`{a} degli studi di Perugia 
and INFN sezione di Perugia, Via A. Pascoli 06100 Perugia, Italy
\\
\email{sergio.scopetta@pg.infn.it} 
}

\date{Received: date / Accepted: date}

\maketitle

\begin{abstract}
The two leading twist, quark helicity conserving
generalized parton distributions (GPDs)
of $^3$He, accessible, for example, in
coherent deeply virtual
Compton scattering (DVCS),
are calculated in impulse approximation (IA).
Their sum,
at low momentum transfer, is found to be largely
dominated by the neutron contribution, so that
$^3$He is very promising
for the extraction of the neutron information. 
Anyway, such an extraction could be  not trivial.
A technique, 
able to take into account the nuclear effects included in the IA analysis
in the extraction procedure, even at moderate values of the
momentum transfer, is  proposed. 
Coherent DVCS arises therefore as a crucial experiment 
to access, for the first time, 
the neutron GPDs and
the orbital angular momentum of the partons in the neutron.
\end{abstract}
\vskip2mm


Generalized Parton Distributions (GPDs) \cite{uno} 
parameterize the non-perturbative hadron structure
in hard exclusive processes, allowing to access unique information
such as, for example,
the parton total angular momentum \cite{rassegne}. 
By
subtracting from the latter the helicity quark contribution, measured in other
hard processes, the parton orbital angular momentum (OAM), contributing to the nucleon
spin, could be then estimated, 
a crucial step towards the solution of
the so called ``Spin Crisis''. 

\begin{figure*}[t]
\vspace{6.5cm}
\includegraphics{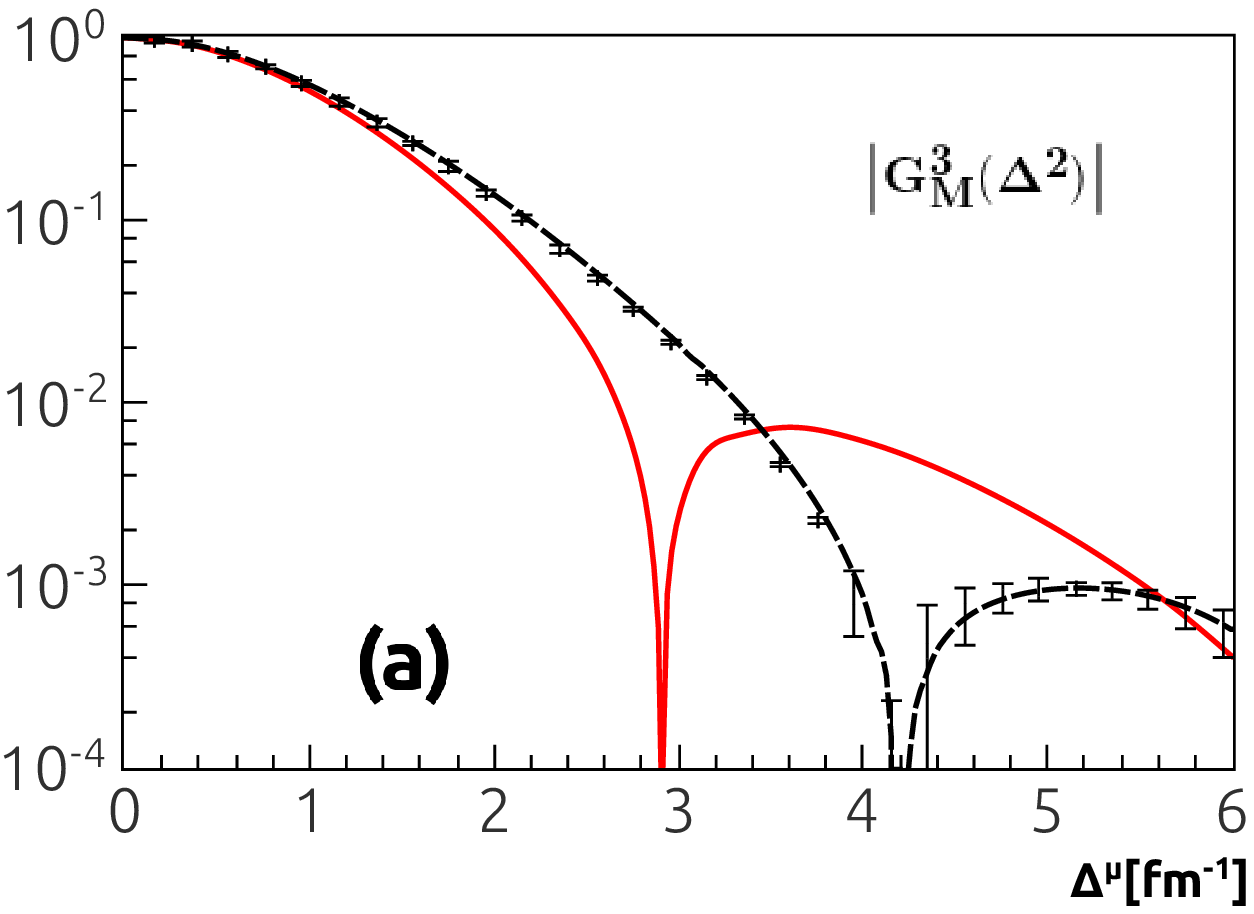}
\end{figure*}

\begin{figure*}[t]
\vspace{6.5cm}
\includegraphics{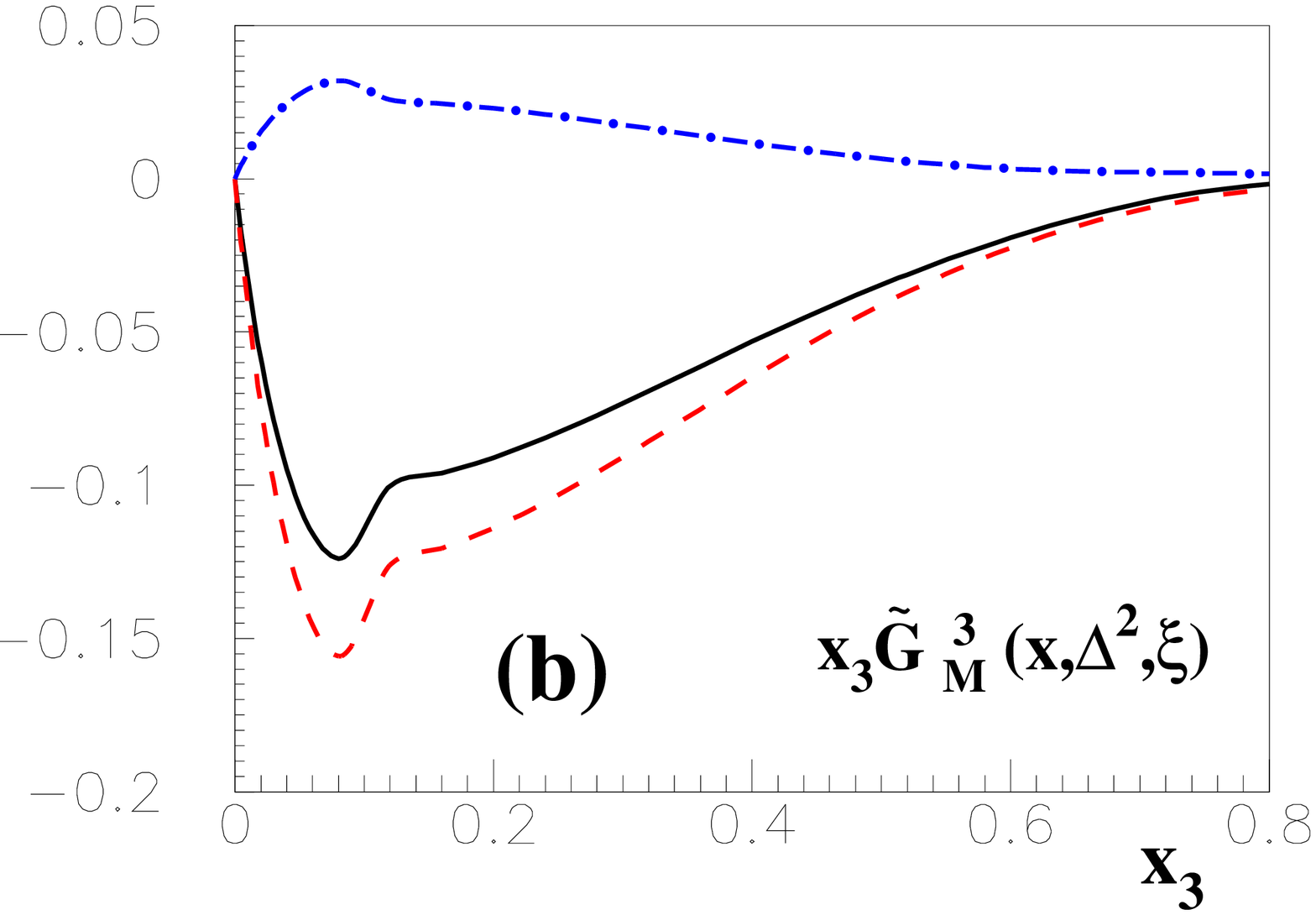}
\vskip-9.3cm
{Fig.1: (a): The magnetic ff of $^3$He, $G_M^3(\Delta^2)$, with $\Delta^{\mu} = \sqrt{-\Delta^2}$. 
Full line: 
the present IA calculation, obtained as the x-integral of $\sum_q \tilde{G}^{3,q}_M$
 (see text). Dashed line: experimental data \cite{dataff}. 
(b): The quantity $x_3 \tilde{G}^3_M(x,\Delta^2,\xi)$, where $x_3 = M_3/M \ x$
and $\xi_3 = M_3/M \ \xi$, shown at $\Delta^2 = -0.1
\ \mbox{GeV}^2$ and $\xi_3=0.1$, \ together with the neutron (dashed) and the proton
(dot-dashed) contribution.}
\end{figure*}

The cleanest process to access GPDs is Deeply Virtual Compton Scattering 
(DVCS), i.e.
$eH \longmapsto e'H' \gamma$ \ when \ $Q^2\gg M^2$ ($Q^2=-q \cdot q$ \ 
is the momentum transfer
between the leptons $e$ and 
$e'$, $\Delta^2$ the one between hadrons $H$ and $H'$ with
momenta $P$ and $P'$, and
$M$ is the nucleon mass.
Another relevant kinematical variable is the so called
skewedness, $\xi = - \Delta^+/(P^+ + P^{'+})$ 
\footnote{In this paper, $a^{\pm}=(a^0 \pm a^3)/\sqrt{2}$}).
Despite severe difficulties to extract GPDs from  experiments, 
data for proton and nuclear targets are being analyzed,
see, i.e., Refs. \cite{data1,data2}.
The measurement of GPDs for nuclei
could be crucial to distinguish
between different models of nuclear
medium modifications of the nucleon structure,
an impossible task in the analysis of DIS experiments only.
Moreover, the neutron measurement, which requires
nuclear targets, is a very relevant information because it permits, 
together with the proton one,
a flavor decomposition of GPDs.
In studies of the neutron polarization, $^3$He plays a special role, 
due its spin structure (see, e.g., Ref. \cite{3He}). 
This is true in particular for GPDs.
In fact, among the latters, the ones of interest here
are $H_q(x,\Delta^2,\xi)$ and
$E_q(x,\Delta^2,\xi)$.
$^3$He, among the light nuclei, is the only one for which the combination
$\tilde{G}_M^{3,q}(x,\Delta^2,\xi) = H^{3}_q(x,\Delta^2,\xi)+
E^{3}_q(x,\Delta^2,\xi)$ of its GPDs
could be dominated by the neutron, being 
$^2$H and $^4$He not suitable to this aim, as discussed
in Ref. \cite{noiold}.
To what extent this fact can be used to extract the neutron information,
is shown in Refs. \cite{noiold,noiarxive}, and summarized here.

The formal treatment of $^3$He GPDs 
in Impulse Approximation (IA) can be found in Refs. \cite{scopetta},
where, for the GPD $H$ of $^3$He,
$H_q^3$, 
a convolution-like equation
in terms of the corresponding
nucleon quantity is found.
Very recently, the treatment has been extended to
$\tilde{G}_M^{3,q}$
(see Refs. \cite{noiold,noiarxive} for details),
yielding

\begin{figure*}[t]

 \vskip-1cm  \begin{minipage}{57mm}

 \hskip -2mm \includegraphics[width=57mm]{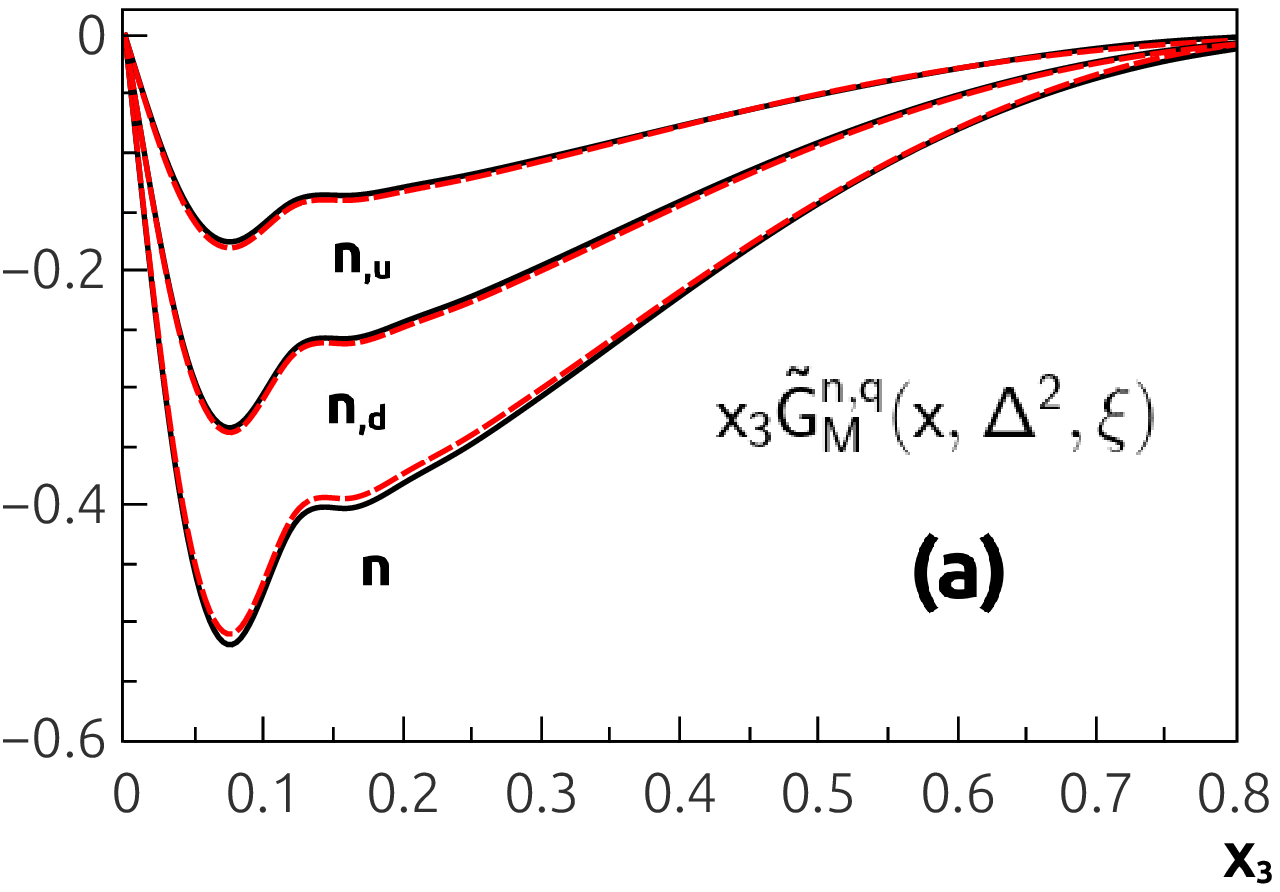}

\end{minipage}

\vskip-4.7cm   \begin{minipage}{57mm}

 \hskip6.7cm 
\includegraphics[width=57mm]{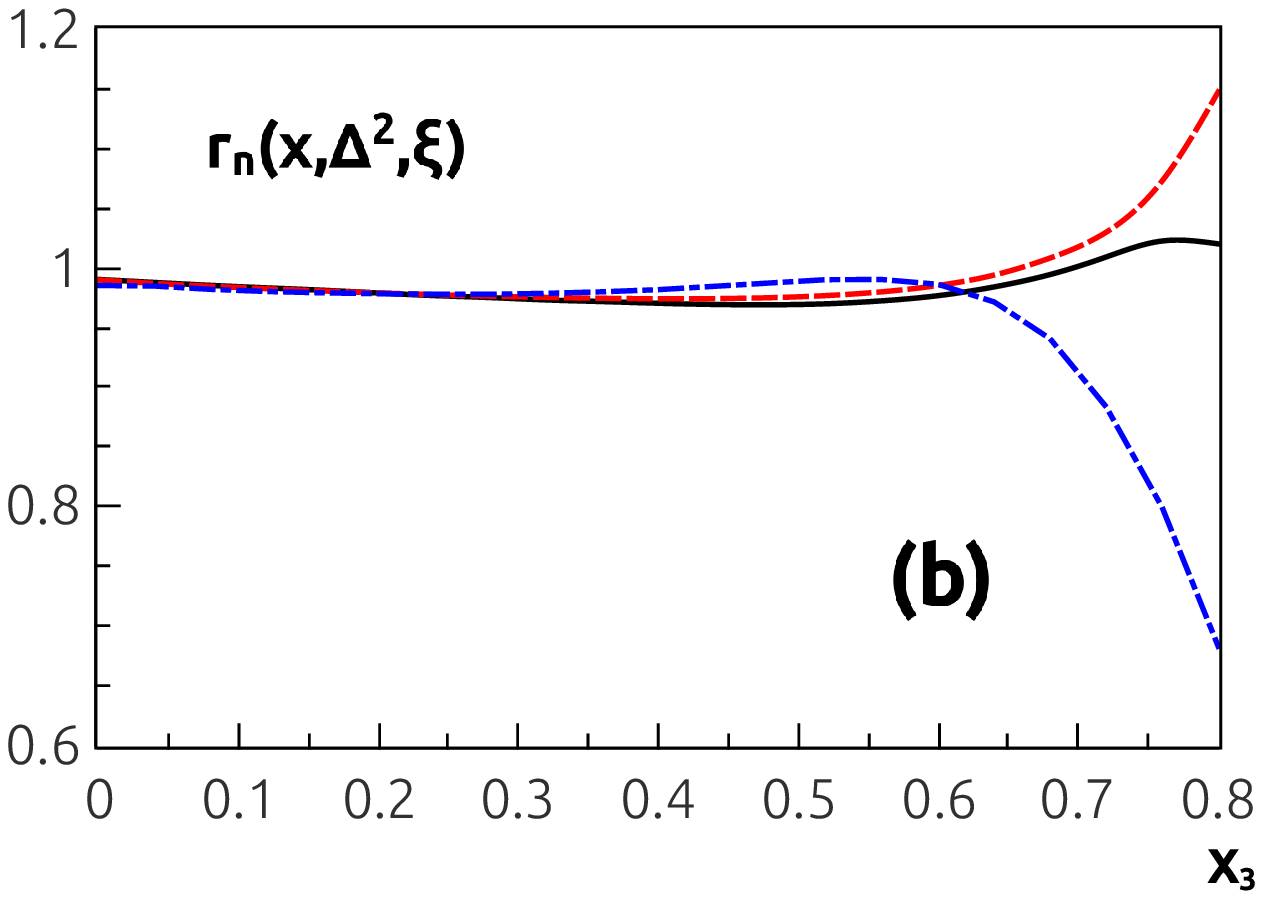}

\end{minipage}

 \vskip 0cm
Fig.2: (a): The quantity $x_3 \tilde{G}^{n,q}_M(x,\Delta^2,\xi)$ for the neutron 
at  
$\Delta^2=-0.1 \ \mbox{GeV}^2$ and $\xi_3=0.1$
 with $u$, $d$ and $u+d$ contributions (full lines), compared with the approximation 
 $x_3 \tilde{G}^{n,q,extr}_M(x,\Delta^2,\xi)$, Eq. (6), (dashed).
(b): The ratio $r_n(x, \Delta^2,\xi)= \tilde{G}_M^{n,extr}(x, \Delta^2,\xi)
/ \tilde{G}_M^{n}(x, \Delta^2,\xi)$, in the forward limit (full), at 
$\Delta^2=-0.1 \ \mbox{GeV}^2$ and $\xi_3=0$ (dashed) and at  
$\Delta^2=-0.1 \ \mbox{GeV}^2$ and $\xi_3=0.1$ (dot-dashed).
\end{figure*}
\vskip-.55cm \begin{eqnarray}
 \tilde G_M^{3,q}(x,\Delta^2,\xi)  = 
\sum_N
\int dE 
\int d\vec{p}~
\tilde{P}^3_N(\vec{p}, \vec{p'},E)
{\xi' \over \xi}
\tilde G_M^{N,q}(x',\Delta^2,\xi'),
\end{eqnarray}
\vskip-3mm
\noindent
where $x'$ and $\xi'$ are the variables for the bound nucleon 
GPDs and $p \, (p'= p + \Delta)$
is its 4-momentum in the initial (final) state. 
Besides, $\tilde{P}^3_N(\vec{p}, \vec{p'},E)$
is a proper combination of components of the spin dependent,
one body off diagonal spectral function:
\newpage
\begin{figure}[t]
\vspace{3cm}
\flushleft
\includegraphics{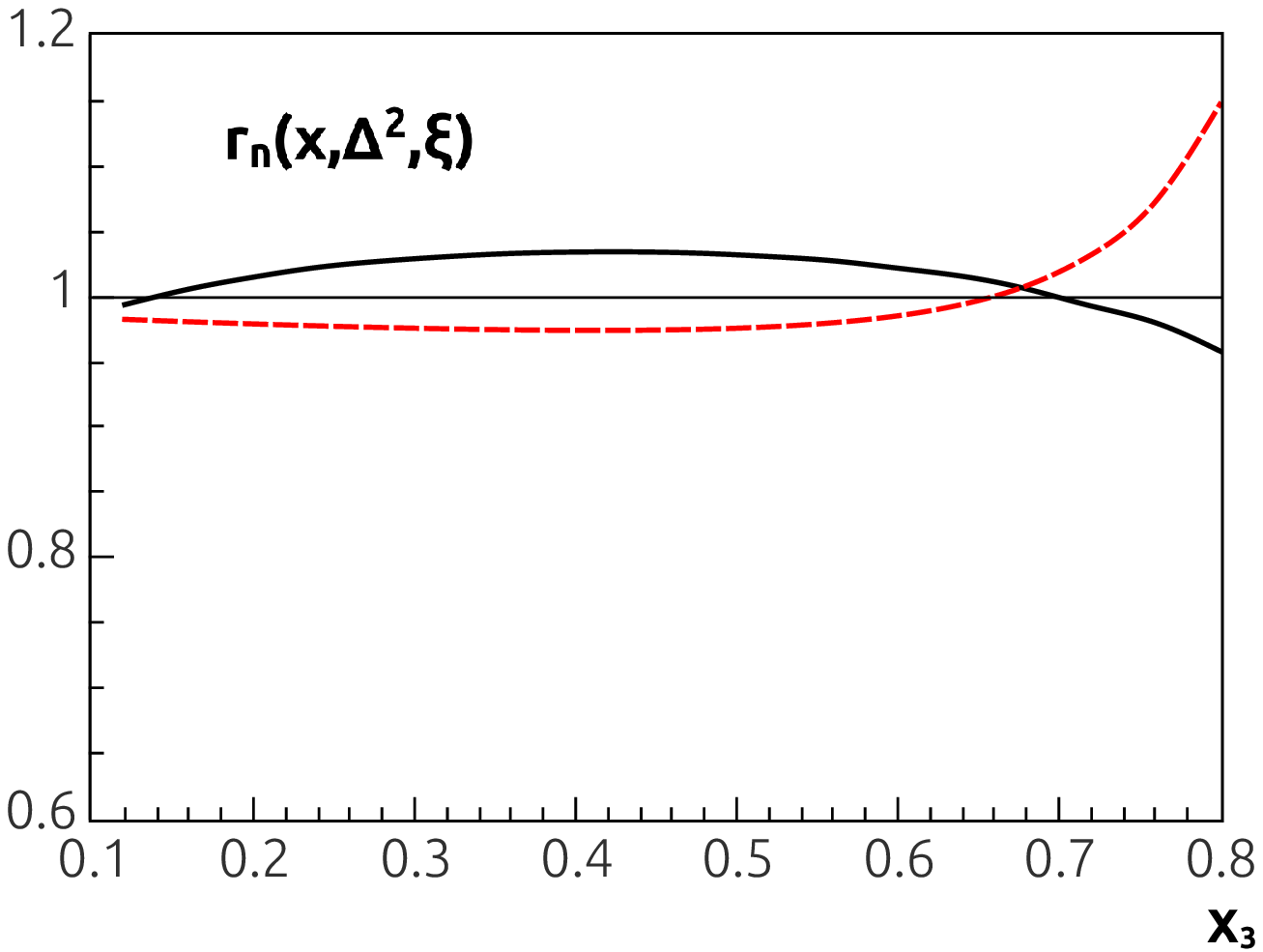}
~~~~~~~~~~~~~~~~~~~~~~~~~~~~~~~~~~~~~~~~~~~~~~~~~~~~~~~~~~~~~~~~~~~~~~
\vskip-3cm
\hskip5.5cm
\begin{minipage}{66mm}
{Fig.3: $r_n(x, \Delta^2,\xi)= \tilde{G}_M^{n,extr}(x, \Delta^2,\xi)
/ \tilde{G}_M^{n}(x, \Delta^2,\xi)$,
at $\Delta^2$ = 0.1 GeV$^2$ and $\xi_3 = 0$, 
using the model of Ref. \cite{Rad1} for the nucleon GPDs (dashed)
and the one of Ref. \cite{sv} (full).}
\end{minipage}
\end{figure}
\ \ \ \ \  \ \ \ \ \ \ \  \ \ \ \ \ \  \ \ \ \ \ \ 

\vskip1.3cm \begin{eqnarray}
 \label{spectral1}
 P^N_{SS',ss'}(\vec{p},\vec{p}\,',E) 
= 
\dfrac{1}{(2 \pi)^6} 
\dfrac{M\sqrt{ME}}{2} 
\int d\Omega _t
\sum_{\substack{s_t}} \langle\vec{P'}S' | 
\vec{p}\,' s',\vec{t}s_t\rangle_N
\langle \vec{p}s,\vec{t}s_t|\vec{P}S\rangle_N~,
\end{eqnarray}
where $S,S'(s,s')$ are the nuclear (nucleon) spin projections
in the initial (final) state, respectively,
and $E= E_{min} +E_R^*$, 
being $E^*_R$ the excitation energy 
of the two-body recoiling system.
The main quantity appearing in the definition
Eq. (\ref{spectral1}) is
the intrinsic overlap integral
\begin{equation}
\langle \vec{p} ~s,\vec{t} ~s_t|\vec{P}S\rangle_N
=
\int d \vec{y} \, e^{i \vec{p} \cdot \vec{y}}
\langle \chi^{s}_N,
\Psi_t^{s_t}(\vec{x}) | \Psi_3^S(\vec{x}, \vec{y})
 \rangle~
\label{trueover}
\end{equation}
between the wave function
of $^3$He,
$\Psi_3^S$,  
with the final state, described by two wave functions: 
{\sl i)}
the
eigenfunction $\Psi_t^{s_t}$, with eigenvalue
$E = E_{min}+E_R^*$, of the state $s_t$ of the intrinsic
Hamiltonian pertaining to the system of two {\sl interacting}
nucleons with relative momentum $\vec{t}$, 
which can be either
a bound 
or a scattering state, and 
{\sl ii)}
the plane wave representing 
the nucleon $N$ in IA.
For a numerical evaluation of Eq. (1),
the overlaps, Eq. (3), appearing in Eq. (2)
and corresponding to the analysis of Ref.
\cite{overlap} in terms of 
Av18  \cite{pot} wave functions
\cite{AV18}, 
have been used, together with a simple
nucleonic model for $\tilde G_M^{N,q}$ \cite{Rad1}
(see Ref. \cite{noiarxive} for details).
%
Since there are no $^3$He data available, 
it is possible to verify only a few general 
GPDs properties, i.e., the forward limit and the first moments.
In particular the calculation of $H^{3}_q(x,\Delta^2,\xi)$ 
fulfills these constraints
\cite{scopetta}. 
In the  $\tilde G_M^{3,q}(x,\Delta^2,\xi)$ case, since there is 
no observable forward limit
for $E^{3}_q(x,\Delta^2,\xi)$, the only possible check is the first moment:
$
\sum_q \int dx \, \tilde G_M^{3,q}(x,\Delta^2,\xi) = G_M^3(\Delta^2);
$
where $G_M^3(\Delta^2)$ is the magnetic form factor (ff) 
of $^3$He.
The result obtained is in perfect agreement with the 
one-body part of the
AV18 calculation
presented in Ref. \cite{schiavilla} (see Fig.1a).
Moreover, for the values of $\Delta^2$ which are relevant for the 
coherent process under investigation here,
i.e., $-\Delta^2 \ll 0.15$ GeV$^2$,
our results compare well also with the data
\cite{dataff}.
With the comfort of this succesfull check, 
results for GPDs of $^3$He are now discussed.
In the forward limit, 
necessary to measure OAM, 
the neutron contribution strongly dominates the
$^3$He quantity, but 
increasing $\Delta^2$ 
the proton contribution grows up
(see Fig.1b), in particular for the $u$ flavor
\cite{noiold,noiarxive}.
It is therefore necessary to introduce
a procedure
to safely extract the neutron information from 
$^3$He
data. This can be done by observing that
Eq. (1) can be written as
\begin{eqnarray}
\tilde G_M^{3,q}(x,\Delta^2,\xi) =   
\sum_N \int_{x_3}^{M_A \over M} { dz \over z}
g_N^3(z, \Delta^2, \xi ) 
\tilde G_M^{N,q} \left( {x \over z},
\Delta^2,
{\xi \over z},
\right)~,
\end{eqnarray}
where $g_N^3(z, \Delta^2, \xi )$ 
is a ``light cone off-forward momentum
distribution'' which,
close to the forward limit, is strongly peaked around $z=1$.
Therefore,
for $x_3 = (M_A/M) x < 1$:

\vskip2.5cm
\vskip-5.5mm
\begin{eqnarray}
\tilde G_M^{3,q}(x,\Delta^2,\xi) 
& \simeq & {low \,\Delta^2} \simeq   
\sum_N 
\tilde G_M^{N,q} \left( x, \Delta^2, {\xi } \right)
\int_0^{M_A \over M} { dz }
g_N^3(z, \Delta^2, \xi ) 
\nonumber
\\
& = &
G^{3,p,point}_M(\Delta^2) 
\tilde G_M^p(x, \Delta^2,\xi) 
+ 
G^{3,n,point}_M(\Delta^2) 
\tilde G_M^n(x,\Delta^2,\xi)~. 
\end{eqnarray}
\vskip-1mm
Here, the magnetic point like ff, 
$G_M^{3,N,point}(\Delta^2)=\int_0^{M_A \over M} dz \, g_N^3(z,\Delta^2,\xi), $ 
which would give the nucleus ff if the proton and the neutron were 
point-like
particles
with their physical magnetic moments, are introduced.
These quantities are very well known theoretically and depend 
weakly on the potential used in the calculation 
\cite{noiarxive}.

\vskip1mm
Eq. (5) can now be used to extract the
neutron contribution:
\vskip-5mm
\begin{eqnarray}
\label{extr}
\tilde G_M^{n,extr}(x, \Delta^2,\xi)  \simeq  
{1 \over G^{3,n,point}_M(\Delta^2)} 
\left\{ \tilde G_M^3(x, \Delta^2,\xi) 
 -  
G^{3,p,point}_{M}(\Delta^2) 
\tilde G_M^p(x, \Delta^2,\xi) \right\}~.
\end{eqnarray}

In Fig. 2a, the comparison between the free neutron GPDs, used 
as input in the calculation,
and  the ones extracted using our calculation for
$\tilde G_M^3$ and the proton model for $\tilde G_M^p$, shows that the
procedure works nicely even beyond the forward limit. The only theoretical
ingredients are the magnetic point like ffs, 
which are completely under control.
This is even clearer in Fig. 2b, where the ratio
$
r_n (x,\Delta^2,\xi)  = 
{\tilde G_M^{n,extr}(x, \Delta^2,\xi)
\over
\tilde G_M^{n}(x, \Delta^2,\xi)}
$
is shown
in a few kinematical regions. The procedure works for $x < 0.7$,
where data are expected from JLab.
Moreover, the extraction procedure depends weakly 
on the used nucleonic model (see Fig. 3 and Ref. \cite{noiarxive}).

In closing, we have shown that coherent DVCS off $^3$He at
low momentum transfer $\Delta^2$ is an ideal process to access 
the neutron GPDs;
if data were taken at higher $\Delta^2$, a relativistic treatment 
\cite{lussino}
and/or the inclusion of many body currents, beyond the present IA scheme, 
should be implemented.

\end{document}